\documentclass[prd,showpacs, 
nofootinbib, preprintnumbers]{revtex4}

\usepackage{amssymb,hyperref, amsmath}
\usepackage[dvips]{color}
\usepackage[dvips]{graphicx}
\usepackage{epsfig}

\preprint{JLAB-THY-09-1059}
\preprint{TCDMATH 09-22}

\begin{document}
\bibliographystyle{apsrev}

\title{An anisotropic preconditioning for the Wilson fermion matrix on the lattice} 

\author{B\'alint Jo\'o}
\email{bjoo@jlab.org}
\affiliation{Thomas Jefferson National Accelerator Facility, Newport News, VA 23606}

\author{Robert G. Edwards}
\email{edwards@jlab.org}
\affiliation{Thomas Jefferson National Accelerator Facility, Newport News, VA 23606}

\author{Michael J. Peardon}
\email{mjp@maths.tcd.ie} 
\affiliation{School of Mathematics, Trinity College Dublin, Ireland} 

\author{for the Hadron Spectrum Collaboration}
\noaffiliation

\date{October 6, 2009}

\begin{abstract}
A preconditioning for the Wilson fermion matrix on the lattice 
is defined, which is particularly suited to the case when the temporal lattice
spacing is much smaller than the spatial one. Details on the implementation of 
the scheme are given. The method is tested in numerical studies of QCD on 
anisotropic lattices.
\end{abstract}

\maketitle


\section{Introduction}
Anisotropic discretisations 
\cite{Morningstar:1997ff,Umeda:2003pj,Morrin:2006tf}
have proved an extremely useful tool for 
determining the energy spectrum of lattice field theories using the Monte Carlo 
method. Four dimensional Euclidean space-time is discretised with separate 
grid spacings $a_s$ and $a_t$ for the spatial dimensions 
and the temporal direction with $\xi=a_s/a_t$. The spectrum of the theory is
determined by measuring the correlation function between operators on the fields
in a Monte Carlo calculation and performing a statistical analysis to determine
the rate of decay of these functions. This fall-off is related to the energies 
of eigenstates of the quantum-mechanical hamiltonian. A fine resolution in the
temporal direction is crucial to make an accurate determination of energies,
particularly when considering massive or highly excited states. The advantage of
the anisotropic lattice is that the rapid rise in numerical effort needed to reduce
the lattice spacing is ameliorated by just reducing the spacing in the 
time direction. The Hadron Spectrum collaboration is currently using these 
lattice regularisations in a large-scale programme 
\cite{Lin:2008pr, Edwards:2008ja, Lin:2007yf}
aiming to measure the QCD
spectrum to high precision. The calculations by the collaboration include the
full dynamics of the lightest three quark fields and in these investigations, 
the dominant source of computational overhead is solving the linear system 
corresponding to quark propagation in a given gauge field background. Quark 
propagation is described by finding solutions to a lattice representation of 
the Euclidean-space Dirac equation:
\begin{equation}
 ( D\!\!\!\!/ + m ) \psi = \eta, 
\end{equation}
where $D\!\!\!\!/$ is the Dirac operator in the presence of the gauge fields. 
The Hadron Spectrum collaboration has used anisotropic lattices with Wilson and
Sheikholeslami-Wohlert \cite{Sheikholeslami:1985ij} discretisations of the 
Dirac operator. 
These solutions, $\psi$ must be
computed during application of the Markov process that generates the ensemble of
gauge fields for Monte Carlo importance sampling as well as during the
measurement phase. The vector space in which the solution is to be constructed
is sufficiently large that direct methods are impractical and instead iterative
solvers must be employed. 
While the anisotropic lattice gives substantial benefits in statistical
precision the disadvantage is that the inclusion of a fine discretisation 
scale; $a_t$; increases the condition number of the linear system to be solved. 
This naturally leads to an increase in the number of iterations needed in the solver. A well-established means of improving efficiency is to precondition 
the problem. 

A quick examination reveals that the dominant terms in the coefficient
matrix of the linear system represent the discretisation of the
temporal derivative in the Dirac operator. In this work, operators
that directly inverts this term are constructed and used to form a
preconditioner. Once the means of applying the inverse of the temporal
term in the Wilson or SW operator has been defined it is seen that
further preconditioning using Schur or ILU factorisation can be
applied. To test this idea, the condition numbers of temporally
preconditioned operators are computed on several gauge field
ensembles. This test uses quenched ensembles since they are
substantially easier to generate with a range of different
anisotropies than fully dynamical ones, and the issue of determining
the parameters in the lattice action needed to give a particular
physical anisotropy $\xi$ is greatly simplified. The temporal
preconditioned matrices to be inverted have condition numbers
approximately twenty times smaller than their unpreconditioned
counterparts and about three to four times smaller than the commonly
used even-odd lattice Schur preconditioning when the anisotropy ranges
from three to six. Taking into account numerical overheads, this
translates to cost reductions of around 1.7-2.4. No advantage is found
using temporal preconditioning in the {\em isotropic} case.

This paper is organized as follows: in section \ref{s:Theory} we set up
our notation, and describe in detail the basics of temporal preconditioning
and how it may be combined with further even-odd preconditioning. In section \ref{s:Numerics} we detail our numerical investigations, including the tuning of our fermion anisotropy parameters and a presentation of condition numbers for various kinds of preconditioning schemes. We sum up and draw our conclusions in section \ref{s:Conclusions}.

\section{Theory}\label{s:Theory}
  The Wilson fermion matrix and its ${\cal O}(a)$-improved
Sheikholeslami-Wohlert (SW) version can be defined on an anisotropic
lattice with three coarse spatial directions with spacing $a_s$ and a
fine temporal spacing $a_t$.  
In practice, the anisotropy is introduced into the simulation by modifying the
couplings in the lattice action that weight temporal and spatial components.
There relative weights are denoted $\gamma_g$ for terms in the gauge action and
$\gamma_f$ for the fermions. These parameters are tuned carefully to ensure
Lorentz invariance is restored in long-distance physics. 
If these parameters are determined using a perturbative expansion, then at the 
tree level they are $\gamma_f = \gamma_g = \xi$. Since practical simulations 
will be performed close to the continuum limit, this relationship should hold 
to within approximately $25\%$. 
Following the notation of \cite{Lin:2008pr},
we will denote the tuned values of these parameters as 
$\gamma^{*}_g$ and $\gamma^{*}_f$ respectively.

With these definitions, the fermion matrix is defined as the sum of the following terms:
\begin{equation}
  M = A + \mu \delta_{x,y} - D_t  - \frac{1}{\gamma_f} D_s, 
\end{equation}
with
\begin{equation}
  D_t = 
        \frac{1-\gamma_0}{2} U_0(x) \delta_{x+\hat{0},y} + 
        \frac{1+\gamma_0}{2} U^\dagger_0(x-\hat{0}) \delta_{x-\hat{0},y}, 
\end{equation}
\begin{equation}
  D_s = \sum_{i=1}^3 \left(
        \frac{1-\gamma_i}{2} U_i(x) \delta_{x+\hat{i},y} + 
        \frac{1+\gamma_i}{2} U^\dagger_i(x-\hat{i}) \delta_{x-\hat{i},y} 
	\right), 
\end{equation}
where $\gamma_i$ for $i=0,1,2,3$ are appropriate spin matrices and $\mu$ is an anisotropic mass term corresponding to bare mass $m_0$
defined as $ \mu = a_t m_0 + 1 + \frac{3}{\gamma_f} $. 

In the case of Wilson fermions, the term $A = 0$
while for the SW improved fermion matrix the mass
term $\mu$ and the hopping terms $D_s$ and $D_t$ are unchanged while
the corresponding expression for $A$ is
\begin{equation}
  A = -\delta_{x,y} \left\{ \frac{c_t}{2} \sum_{i=1}^3 \sigma_{i0}
      F_{i0}(x) + \frac{c_s}{2} \sum_{ij} \sigma_{ij} F_{ij}(x) \right\} \ ,
\end{equation}
where $\sigma_{ij} = \frac{1}{2}\left[\gamma_{i}, \gamma_j \right]$ is a
commutator of the spin matrices and $F_{ij}(x)$ is the
anti-hermitian ``clover-leaf'' term constructed from the plaquettes in the 
$i,j$ plane
emanating from from site $x$ as defined
in \cite{Lin:2008pr} and elsewhere.

The tree level tadpole improvement coefficients $c_s$ and $c_t$ are
defined as 
\begin{equation} \label{eq:ClovCoeffs}
c_s = \frac{1}{u^3_s} \frac{1}{\gamma_f}, \qquad c_t = 
\frac{1}{2 u^2_s u_t}\left( \frac{\gamma_g}{\gamma_f} + \frac{1}{\xi} \right), 
\end{equation}
where $u_s$ is the spatial tadpole factor computed from the spatial
plaquettes of our gauge ensembles $U_{ss}$ as
\begin{equation}
u_s = \langle \frac{1}{3}U_{ss} \rangle^{\frac{1}{4}}, 
\end{equation}
and we have set $u_t=1$. The isotropic case is recovered when 
one sets $u_s=u_t$, $\gamma_g=\gamma_f=1$ and
so $c_s = c_t$ are the same tree level improved tadpole coefficient.

As in the isotropic case, these linear operators act on vectors in an
$N_c \times N_\gamma \times N_s$-dimensional space of complex fermion
fields, where $N_c$ is the number of colors in the gauge group,
$N_\gamma=4$ is the number of components in a spin-1/2 representation
of the group of four-dimensional Euclidean rotations and $N_s$ the
number of sites on the four-dimensional lattice.

The additional terms in the SW action are all diagonal in the spatial
and temporal lattice co-ordinates.  With the matrix expressed in this
form, the largest terms in $M$ are contained in $A$, $D_t$ and $\mu$,
and the contribution from terms in $D_s$ is smaller by a factor of the
(bare fermion) anisotropy. This suggests an efficient preconditioning
for this matrix can be constructed if a way of applying the inverse of
$A + \mu - D_t$ is known. Notice this matrix is block-diagonal in the
spatial lattice co-ordinates, so solving this problem requires being
able to invert a one-dimensional lattice operator.

  \subsection{A temporal preconditioner}
    We now proceed to present the technique of the temporal
preconditioning. In the discussion below, we will initially focus on
the Wilson case ($A=0$) for simplicity. We will then comment upon the
strategies available to deal with the full SW case.
We begin by defining temporal spin-projection operators 
\begin{equation}
   P_{\pm} = \frac{1}{2} (1\pm \gamma_0), 
\end{equation}
and a temporal hopping matrix at every spatial site on the lattice
\begin{equation}
   T_{t,t'}(\vec{x}) = \mu \delta_{t,t'} - 
             U_0(\vec{x},t) \delta_{t+1,t'}, \quad t=0 \hdots N_t -1, 
\end{equation}
where $N_{t}$ is the number of lattice points in the temporal direction 
of the lattice. The operator $T$  is gauge covariant with no spin structure. 
Further, $T$ can have a variety of boundary conditions. We will consider
primarily the cases where $T$ is either periodic or anti--periodic where 
\begin{eqnarray}
  T_{N_t-1,N_t-1}(\vec{x}) &=& \mu \\
  T_{0,N_t-1}(\vec{x})  &=& \pm U_{0}(\vec{x},N_t-1) \ , \label{eq:boundary} 
\end{eqnarray}
and the sign in Eqn.~\ref{eq:boundary} is positive or negative when the boundary conditions are respectively periodic or anti-periodic.

We now define operators on all spin components,
\begin{equation}
   C_L^{-1} = P_+ + P_- T \mbox{ and } 
   C_R^{-1} = P_- + P_+ T^\dagger, 
\end{equation}
where the invertability of these matrices has been assumed. 
With these definitions, the temporal hopping term can be expressed as
\begin{equation}
  \mu - D_t = C_L^{-1} C_R^{-1}  \ .
    \label{eqn:dt_clcr} 
\end{equation}
Constructing $C_L^{-1}$ and $C_R^{-1}$ with the spin-projector structure given 
above ensures they obey the relation
\begin{equation}
  C_L^{\dagger-1} = \gamma_5 C_R^{-1} \gamma_5, 
\end{equation}
which will allow the construction of a preconditioned matrix that maintains
$\gamma_5$ hermiticity. Eqn.~\ref{eqn:dt_clcr} now shows how $C_L$ and $C_R$ 
will make a useful preconditioner for the Wilson fermion matrix on the
anisotropic lattice. 
The fermion matrix can be written as
\begin{equation}
   M = C_L^{-1} C^{ }_L M C^{}_R C_R^{-1} = C_L^{-1} \tilde{M} C_R^{-1}, 
\end{equation}
with $\tilde{M} = C^{ }_L M C^{}_R$. For Wilson fermions,
\begin{equation}
   \tilde{M} = I - \frac{1}{\gamma_f}C_L D_s C_R, 
\end{equation}
and the preconditioned matrix is equal to the identity plus terms
proportional to $1/\gamma_f$ only.


The operation of $\tilde{M}$ on a fermion field requires the operation
of $C_L=(P_+ + P_- T)^{-1}$. Since $P_+$ and $P_-$ define orthogonal
projectors, this inverse can be re-written as $C_L = P_+ + P_- T^{-1}$
and the application of $C_L$ is reduced to finding the inverse of
$T$. If the lattice fields had Dirichlet boundary conditions, $T$ (a
lattice representation of the forward-difference operator) could be
inverted easily by back-substitution, starting at the open
boundary. Most lattice calculations use periodic or anti-periodic
boundary conditions however and so back-substitution is
insufficient. The Sherman-Morrison-Woodbury
(SMW) \cite{Sherman:1949,Sherman:1950,Woodbury:1950} formula
provides a simple means of inverting a matrix that differs in a small
number of elements from another matrix whose inverse is
computationally cheap to apply. The forward difference operator with
periodic boundary data, $T$ can be written in terms of the difference
operator with open boundaries, $T_0$ and a correction term:
\begin{equation}
   T = T_0 + V W^\dagger, 
      \label{eqn:SMW1}
\end{equation}
with 
\begin{equation}
   [T_0]_{t,t'}(\vec{x}) = 
   \left\{
      \begin{array}{ll} 
   \mu \delta_{t,t'} - U_0(\vec{x},t) \delta_{t+1,t'} 
         & \mbox{ if }\;\; t=0..N_t-2 \\
   \mu \delta_{t,t'}  
         & \mbox{ if }\;\; t=N_t-1, 
      \end{array}\right.
\end{equation}
and $V$ and $W$ are $N_c N_t \times N_c$ column matrices where the only non-zero
entries are in either the first or last sites:
\begin{equation}
   V_{t}(\vec{x}) = - U_0(\vec{x},t) \delta_{t,N_t-1}, 
\end{equation}
and 
\begin{equation}
   W_{t}(\vec{x}) = \delta_{t,0}.
\end{equation}
Since $T$, $T_0$, $V$ and $X$ are
defined on each spatial site $\vec{x}$, we will suppress the spatial
index in subsequent discussion except where it may be needed for
clarity. With these definitions, $V W^\dagger$ is a rank $N_c$
correction that adds the effects of the boundary condition back in to
the open-boundary-data operator.  The SMW formula then gives an
expression for the inverse of $T$ defined in
Eqn.~\ref{eqn:SMW1}. Defining $X = T_0^{-1} V$ yields
\begin{equation}
   T^{-1} = T_0^{-1} - X (I + W^\dagger X)^{-1} W^\dagger.
\end{equation}
All that remains to be evaluated is $(I + W^\dagger X)^{-1}$ but note that this
is a small (rank $N_c$) matrix at each spatial site whose inverse is 
straightforwardly computed. In practice, the algorithm proceeds as follows:
\begin{enumerate}
  \item Prior to use, the preconditioner is initialized. On each spatial site $\vec{x}$,  the expression 
    \begin{enumerate} 
      \item $X = T_0^{-1} V$ is computed by back-substitution and
      \item $\Lambda = (I + W^\dagger X)^{-1}$ is subsequently 
      computed. 
    \end{enumerate} 
    All these results are stored. $\Lambda$ requires storage of just
    $N_c \times N_c$ complex numbers per spatial site, while $X$
    requires $N_c \times N_c N_t$. This is smaller storage requirement
    than a single fermion field. Also, we note that one can
    immediately compute $X \Lambda$ at this point which also requires
    storage of $N_c \times N_c N_t$ per spatial site but which may
    overwrite the original $X$.
  \item Given a particular right-hand side $\eta$, computing 
  $\psi = T^{-1} \eta$ requires first evaluating
  \begin{enumerate}
    \item $\chi = T_0^{-1} \eta$ by back-substitution, then
    \item $q = \Lambda W^\dagger \chi$. Note that $W^\dagger \chi$ is just the
    $N_c$-component vector on time-slice $t=0$ of vector $\chi$ and so
    evaluation of $W^\dagger \chi$ is computationally trivial. Finally, the
    solution is formed:
    \item $\psi = \chi - X q$
  \end{enumerate}
\end{enumerate}

The back substitution process for $X$ can formally 
be carried out analytically. For the case of $T$ one has:
\begin{eqnarray}
X(\vec{x}, N_t-1) &=& - \frac{1}{\mu}U_{t}(\vec{x}, N_t-1),\\
X(\vec{x}, t ) &=& -\frac{1}{\mu^{N_t-t}}\prod_{j=N_t-t}^{N_t-1}U_{t}(\vec{x},j)
\ \mbox{for $0 \le t < N_t -1$},
\end{eqnarray}
and successive terms in $X$ are suppressed by powers of
the mass term $\mu$, and the matrix product forms a series
which for $X(\vec{x},0)$ is the Polyakov loop.  In principle; for
large enough $N_t$; one could find some $k$ such that for $0 \le t <
k$ one has $X(\vec{x},i) \approx 0$ numerically, and one may then save some
numerical effort by just setting those values of $X(\vec{x},t)=0$ and not
evaluating matrix products using them but setting them to zero
also. Since the link matrices in the components of $X$  are
$SU(3)$ one knows that their product is also $SU(3)$ and hence one can
find the norm of $X(\vec{x},i)$ as
\begin{equation}\label{eq:cutoff}
|| X(\vec{x},t) || = \sqrt{3} \mu^{-\left( N_t - t \right) }, 
\end{equation}
where the factor of $\sqrt{3}$ comes from the $SU(3)$ nature of the
link matrices. We will refer to the cutting off the computation of
$X(\vec{x},t)$ for sufficiently small values of $t$ as the {\em cutoff
  trick}. 
Caution should be used however to ensure there is no impact in the
precision of final result.

The inverse of $T^\dagger$, required for the operation of $C_R$ is
formed in the same way, using appropriate redefinitions of $X, W$ and
$V$ and with forward-substitution solves. We note that in step 1(b)
above, only an $N_c \times N_c$ complex matrix needs to be inverted per
site. The cutoff trick also works, but now the terms are least
suppressed at $t=0$ (open end of the forward substitution) and most
suppressed at $t=N_t-1$, building up to the Hermitean conjugate of the
Polyakov loop for $X(\vec{x},N_t-1)$,
and the  $N_t-t$ term in eqn. \ref{eq:cutoff} needs to be
replaced with $t$.

Let us now comment on the case for SW fermions. Proceeding as above, the entire procedure is valid, but the preconditioned matrix changes to:
\begin{equation}\label{eq:ContaminatedUnprec}
\tilde{M} = C_L M C_R = I + C_L \left( A - \frac{1}{\gamma_f} D_s \right) C_R \ .
\end{equation}

It then becomes tempting to extend the definition of 
$C^{-1}_L$ and $C^{-1}_R$ in such a way that
\begin{equation}
 A + \mu - D_t = C^{-1}_L C^{-1}_R, 
\end{equation}
so that the preconditioned matrix would maintain its original $I -\frac{1}{\gamma_f} C_L D_s C_R$ form, even in the case of a general, non-zero SW term, \i.e. $A \neq 0$.  The practical difficulty with this approach is that the
$\sigma_{\mu\nu}$ terms in $A$ couple all spin components and the
forward and backward difference operators in $D_t$ cease to be
directly separable. Correspondingly the construction of suitable $C_L$
and $C_R$ terms would require the inversion of a block-tridiagonal
matrix, with $N_\gamma N_c \times N_\gamma N_c$ sized blocks rather
than just $N_c \times N_c$. Further, the back/forward substitutions
fill-in these blocks destroying the site wise block-diagonal structure
of the SW matrix and so, the inversions of the diagonal blocks would
need dense inversions of the full $N_\gamma N_c$ dimensional
sub-blocks. We will refer to this approach as {\em full SW temporal preconditioning}. The details of this approach are discussed in the appendix.

Nonetheless, even if one just uses the same $C_L$ and $C_R$ as for the
Wilson action and suffers the contamination from the SW term in the
preconditioned matrix $\tilde{M}$ in
eq. (\ref{eq:ContaminatedUnprec}), it can be seen that the $D_t$ term
is still inverted, and that the $D_s$ terms are suppressed by a factor
of $\frac{1}{\gamma_f}$.  Hence one can expect that this form of preconditioning
is still more effective than using the unpreconditioned operator. In 
what follows we will refer to this approach of using the $C_L$ and $C_R$ preconditioners from the case of the Wilson fermion matrix to preconditione the SW operator, as {\em partial SW temporal preconditioning}.

  \subsection{Combining the temporal preconditioner with other schemes}
%
The usual isotropic Wilson and SW operators are efficiently preconditioned by
considering a Schur decomposition after first ordering lattice sites according
to their four-dimensional co-ordinate parity, 
$p_4(x) = (-1)^{x_0 + x_1 + x_2 +x_3}$. One has
\begin{equation}
  M = \left(\begin{array}{cc} M^{ee} & M^{eo} \\
                              M^{oe} & M^{oo} \end{array} \right)
  = 
  \left(\begin{array}{cc} I^{ee} & 0 \\
                              M^{oe} [M^{ee}]^{-1} & I^{oo} \end{array} \right)
  \left(\begin{array}{cc} M^{ee} & 0 \\
                              0 & M^{oo} - M^{oe} [M^{ee}]^{-1} M^{eo} 
			        \end{array} \right)
  \left(\begin{array}{cc} I^{ee} & [M^{ee}]^{-1} M^{eo} \\
                              0 & M^{oo} \end{array} \right), 
\end{equation}
with $M^{ee(oo)} = A^{ee(oo)}+\mu$, and $M^{eo} = -D^{eo}_w$, $D^{eo}_w= \frac{1}{\gamma_f} D_s + D_t$ being the 4-dimensional Wilson Dslash
operator, and the {\em Schur preconditioned} matrix is
\begin{equation}
\tilde{M} =\left(\begin{array}{cc} M^{ee} & 0 \\
                              0 & M^{oo} - M^{oe} [M^{ee}]^{-1} M^{eo} 
			        \end{array} \right) \ .
\end{equation}
This kind of Schur preconditioning with a $p_4$ ordering,  is the standard even-odd preconditioning method in use for Wilson and SW fermions, and we shall refer to it as {\em 4D-Schur preconditioning} from now on.

This idea can be combined with the temporal preconditioner with a
small modification; instead of ordering lattice sites by a
four-dimensional parity, a three-dimensional equivalent is used,
$p_3(x) = (-1)^{x_1 + x_2 +x_3}$.  The preconditioned matrix
$\tilde{M}$ retains its form in terms of $M^{ee(oo)}$ and
$M^{eo(oe)}$, however these now change their meaning slightly, since
with this ordering, the operators $A$ and $D_t$ connect lattice sites
with the same $p_3$ while the spatial hopping matrix couples sites
with opposite $p_3$. and
\begin{equation}
M^{ee(oo) } = A^{ee(oo)} + \mu - D_t^{ee(oo)} \ , \qquad M^{eo(oe)} =
-\frac{1}{\gamma_f}D^{eo(oe)}_s.
\end{equation}

For the Wilson action, the inverse of the block matrix, $M^{ee} = \mu
- D_t^{ee}$ is formed using the method described in the previous
section.  For full temporal preconditioning in the clover case, one
would need to form the inverse of $M^{ee} = A^{ee} + \mu -D_t^{ee}$
the difficulties with which have already been discussed, and we can
refer to the result as {\em temporal preconditioning combined with
  3D-Schur preconditioning.}  Alternatively, one can proceed with
partial preconditioning for the Clover case, using the $p_3$ ordering
preconditioner appropriate for the Wilson action. In this case we can
refer to the result as {\em temporal preconditioning combined with 3D incomplete lower-upper (ILU) preconditioning}.

First, we define the action of the left temporal preconditioner on the even 
3-parity sub-lattice to be $C^{e}_L$ and define correspondingly the right
preconditioner and both their odd sub-lattice counterparts to be 
$C^{e}_R, C^{o}_L, C^{o}_R$. We also introduce the notation that for
some generic term $K$ the corresponding term $\bar{K}$ is defined as $\bar{K} = C_L K C_R$. Hence,
\begin{equation}
   \bar{D}_s^{eo} = C_L^{e} D_s^{eo} C_R^{o}.
\end{equation}

To keep the discussion below general, we will also define the operator
\begin{equation}
\bar{Q} = C_L \left( A + \mu - D_t \right) C_R \ .
\end{equation}
In the case of Wilson fermions, $A=0$ and so $\bar{Q}=I$. 
For SW fermions with full SW preconditioning one also has $\bar{Q}=1$ while 
with partial SW preconditioning, 
\begin{equation}
\bar{Q} = 1 + C_L A C_R = 1 + \bar{A} \ .
\end{equation}
$\bar{Q}$ is diagonal in terms of $p_3$ even-odd 
indices,
and we may refer to its even-even (odd-odd) sub blocks $\bar{Q}^{ee(oo)}$
using $C^{e(o)}_L$, $C^{e(o)}$ and $A^{ee(oo)}$ as needed.

With these expressions, the even-odd preconditioners for the 
Wilson matrix become
\begin{equation}
  S_L = \left(\begin{array}{cc} C_L^{e} & 0 \\
         \frac{1}{\gamma_f} \bar{D}_s^{oe} C_L^{e} & C_L^{o} \end{array} \right)
  \mbox{ and } 
  S_R = \left(\begin{array}{cc} C_R^{e} & \frac{1}{\gamma_f} C_R^{e} \bar{D}_s^{eo}
     \\
         0 & C_R^{o} \end{array} \right), 
\end{equation}
which gives
\begin{equation}
  \tilde{M}_{3} = S_L M S_R = \left(\begin{array}{cc} \bar{Q}^{ee} & -\frac{1}{\gamma_f}\left[ 1 - \bar{Q}^{ee} \right] \bar{D}_s^{eo} \\
 -\frac{1}{\gamma_f} \bar{D}_s^{oe} \left[ 1-\bar{Q}^{ee} \right]  & \bar{Q}^{oo} - \frac{1}{\gamma^2_f} 
			     \bar{D}_s^{oe} \left[ 2 - \bar{Q}^{ee} \right]  \bar{D}_s^{eo} 
			     \end{array} \right). 
\end{equation}

If $\bar{Q}=1$, as is the case of Wilson fermions or Clover fermions with full 
temporal preconditioning, this matrix reduces to
\begin{equation} \label{eq:TPrecFull}
  \tilde{M}_{3} = \left(\begin{array}{cc} 1 & 0 \\
 0  & 1 - \frac{1}{\gamma^2_f} 
			     \bar{D}_s^{oe} \bar{D}_s^{eo} 
			     \end{array} \right),  
\end{equation}
so that the  preconditioned matrix differs from the identity only by terms proportional to $\frac{1}{\gamma^2_f}$. 
In the case of partial SW preconditioning, where $\bar{Q}=1+\bar{A}$,
the preconditioned matrix is
\begin{equation}
  \tilde{M}_{3} = \left(\begin{array}{cc} 1+\bar{A}^{ee} & \frac{1}{\gamma_f} \bar{A}^{ee} \bar{D}_s^{eo} \\
 \frac{1}{\gamma_f} \bar{D}_s^{oe} \bar{A}^{ee}  & 1 + \bar{A}^{oo} - \frac{1}{\gamma^2_f} 
			     \bar{D}_s^{oe} \left[ 1 - \bar{A}^{ee} \right]  \bar{D}_s^{eo} 
			     \end{array} \right).  \label{eq:TPrecPartial}
\end{equation}
We can see from eq. \ref{eq:TPrecPartial} that in contrast to full preconditioning in eq. \ref{eq:TPrecFull}, we now have non--zero off diagonal elements (in even-odd space) that are only suppressed by $\gamma_f$. Further the diagonal elements contain components proportional to $a_t$ in the Clover terms $\bar{A}$. These terms can counter potential suppression by $\gamma_f^2$ in the odd-odd checkerboarded term of $\tilde{M}_3$.

To complete this discussion, we note that effective use of the preconditioner
requires the inverses $S^{-1}_L$ and $S^{-1}_R$ to be applied so the 
solutions with the unpreconditioned matrices may be obtained. Using the definitions of $C^{-1}_{L(R)}$ restricted to the even and odd sites respectively we have
\begin{equation}
  S^{-1}_L = \left(\begin{array}{cc} \left( C_L^{e} \right)^{-1} & -\frac{1}{\gamma_f}\bar{D}^{eo} \left( C_L^{o} \right)^{-1} \\
        0  & \left( C_L^{o} \right)^{-1} \end{array} \right)
  \mbox{ and } 
  S^{-1}_R = \left(\begin{array}{cc} \left( C_R^{e} \right)^{-1} & 0 \\
 -\frac{1}{\gamma_f} \left( C_R^{e} \right)^{-1} \bar{D}_s^{eo} & \left( C_R^{o} \right)^{-1} \end{array} \right) \ .
\end{equation}

  \subsection{Numerical cost of ILU Scheme} \label{s:FLOPCost}
    \newcommand{\C}{\ensuremath \mathcal{C}}

We consider the partial temporally preconditioned scheme, combined with ILU even-odd preconditioning to be potentially the most attractive, since it is simple in terms of implementation and is equivalent to the full 3D Schur preconditioned scheme in the case of Wilson fermions. However, applying the preconditioners does incur some numerical overhead. The overhead depends to some degree on details of the implementation of the method. We will consider two implementations below.

First we consider the naive implementation of the method, with no gauge fixing and in spinor basis where $\gamma_0$ is not diagonal. This could be the case in a general code, using a chiral spin-basis such as the Chroma software system \cite{Edwards:2004sx}.
 Neglecting the cost of preparing the sources and recovering the
 solutions (using $S^{-1}_L$ and $S^{-1}_R$) we can compare costs of
 the usual 4D Schur preconditioning and the partially temporally
 preconditioned ILU scheme, which we will denote as $\C(\tilde{M}_{\rm
   4D Schur})$ and $\C(\tilde{M}_{\rm ILU})$ respectively, by counting
 the floating point operations (FLOPs) in the respective
 preconditioned linear operators.

Relegating the actual counting of FLOPs to appendix \ref{s:Count}, we
merely state here that the ratio of floating point costs:
\begin{equation}
R = \frac{ \C(\tilde{M}_{\rm ILU}) }{ \C(\tilde{M}_{\rm 4D Schur})}, 
\end{equation} 
is
\begin{equation}
R^{W}  =  \frac{ 3432 N_t - 576}{2668 N_t} \approx 1.286-\frac{0.216}{N_t},
\end{equation}
for Wilson fermions, and 
\begin{equation}
R^{C} =  \frac{ 6012 N_t - 1152 }{3732 N_t } \approx 1.611 - \frac{0.309}{N_t},
\end{equation}
for Clover fermions respectively, resulting typically in about a 29\%
overhead for Wilson, and 61\% overhead for Clover from the ILU scheme in 
terms of FLOPs as compared to the standard 4D Schur even-odd scheme.
This must be matched by the gain in terms of condition
number from the preconditioner  for it to remain competitive.

Concurrent with writing this paper, some clever optimization techniques were brought to our attention by the authors of \cite{MikeGPU,MikeGPU2} arising from work with General Purpose Graphics Processing Units (GPGPUs). These techniques save both memory bandwidth and FLOPs. 
The first technique we consider is to fix the gauge prior to the
inversion proess, to the Axial gauge (temporal gauge). The effect of
this operation is that all the links that are not on the temporal
boundary are transformed to the unit matrix: $U_{0}(\vec{x},t)=1$ for
$t=0\hdots N_t-2$. This can save floating point operations in the
back(forward) substitutions with $T_0$, where in each step one can
save an SU(3) matrix-color vector multiply. Further, one can save on
memory requirements since the block vector $X$ is simplified to
\begin{equation}
X(\vec{x},t)=\frac{1}{\mu^{N_t-t}}U_{0}(\vec{x}, N_t -1), 
\end{equation}
with the other terms in the partial Polyakov loops now being the identity.
Correspondingly, instead of storing all of $X$, one can easily compute
any component of it from the boundary link matrix which one stores anyway.
Fixing to the axial gauge is a straightforward operation which can be
amortized over either one and especially over several solves.

The second trick that can prove useful is to employ the Dirac-Pauli 
spin basis in which $\gamma_0$ is diagonal. This simplifies the 
projector operators $P_{+}$ and $P_{-}$ so that they select the top or
bottom two spin components of four spinors respectively. This can save
FLOPs on spinor reconstuction (which now no longer needs to be done 
in the time direction) but also when one has a sum of the form 
\begin{equation}
S = P_{+} \psi + P_{-} \chi, 
\end{equation}
one has no arithmetic to perform, since the spin components filtered
by the projectors can be written directly into the correct components
of $S$ without requiring any addition.
We enumerate in explicit detail the savings from these two implementation
techniques in appendix \ref{s:CountWithTricks}. It is shown there that
employing both of these techniques can save roughly 22-25\% in terms
of FLOPs over the naive implementation.

When compared to the 4D-Schur preconditioned scheme, which also benefits
from these improvements, we find that the techniques result in relative
overhead ratios of:
\begin{equation}
R^{W}  =  \frac{ 2664 N_t - 48}{2044 N_t} \approx 1.30-\frac{0.023}{N_t}, 
\end{equation}
for Wilson fermions, and 
\begin{equation}
R^{C} =  \frac{ 4476 N_t - 96 }{3108 N_t } \approx 1.440 - \frac{0.031}{N_t},
\end{equation}
for Clover fermions. In particular, the relative overhead for the
Clover operator appears to be substantially reduced compared to the naive
implementation (44\% as opposed to the previous 61\%).
The foregoing discussion does not make use of the cutoff trick, which
can be used to further reduce the floating point costs of the
preconditioned operators as discussed earlier.

\section{Numerical Investigation} \label{s:Numerics}
  \subsection{Strategy and Choice of Parameters}\label{s:NumStrategy}
Wishing to investigate the efficacy of the preconditioning strategies
as functions of both lattice anisotropy and quark mass in as
realistic a setting as possible, we have opted to measure the
condition numbers of the various operators in three quenched
ensembles. These ensembles were chosen to have target (renormalized)
anisotropies of $\xi=1.0$, $\xi=3.0$ and $\xi=6.0$ respectively, thus
ranging from the fully isotropic to the highly anisotropic. We picked
a wide range of quark masses, to give pion masses in the range of
about $450$ MeV to $750$ MeV. We note that this necessitated us tuning
the fermion anisotropies $\gamma_f$ so as to make the renormalized
anisotropies $\xi$ (as fixed by the pion dispersion relation) the
same as our target anisotropies $\xi$, a subject we will discuss in
more detail further on.

During our study, we have opted to keep the physical temporal extent
of the lattice fixed in the time direction. This approach means
that as we increased the anisotropy, we likewise increased the number of
lattice points in the time direction. We have used $N_t=16$, $N_t=48$,
$N_t=96$ for the anisotropies of $\xi=1.0$, $\xi=3$ and $\xi=6$
respectively. This increase of the temporal resolution may have an
effect on the condition numbers of our operators. We felt however,
that this is typically the approach one would use in a real
calculation, rather than keeping the temporal extent fixed, and that
we should absorb this effect in our efficiency estimates, in order to
give a realistic measure of the performance of the preconditioning.

\subsection{Code and Computers}
We coded the temporal preconditioned Wilson-Clover operators, in the
Chroma \cite{Edwards:2004sx} software system. We implemented both 3D
ILU and 3D Schur even-odd preconditionings in space, in combination
with the temporal preconditioning. In the case of the 3D Schur even-odd
preconditioning, we used an inner Conjugate Gradients solve, to invert
$M^{ee}$ in the Schur complement. We also used the unpreconditioned
and 4D Schur Even-Odd preconditioned operators already present in the
Chroma suite to measure reference results. Our implementations used
the naive implementation technique discussed in
sec. \ref{s:FLOPCost}.  In order to tune the fermion anisotropies, we
carried out some hadron spectroscopy calculations, in particular
the measurement of the pseudoscalar correlation functions at various
momenta. In order to measure the condition numbers of the square
operators, we used the Ritz-minimization technique of
\cite{Bunk:1996kt}.  Both these sets of measurements are standard
within the Chroma distribution.  Fitting of our spectroscopy results
used the so called 4H code developed by UKQCD.  Our calculations were
carried out on the Jefferson Lab 6n and 7n clusters.

\subsection{Selecting the Gauge Action Parameters}

We chose our isotropic reference case, to be a quenched dataset with
the Wilson gauge action at $\beta=6.0$ as it is well known in the
literature to have  a lattice spacing of $0.1$fm
\cite{Michael:1995bi}. In the anisotropic case using $\beta=6.1$ has
approximately the same spatial lattice spacing at $\xi=3$ and
$\xi=6$ \cite{Klassen:1998ua}. We chose the bare gauge anisotropies
from the formula suggested by \cite{Klassen:1998ua}. Our
gauge production parameters are summarized in Table~\ref{tab:GaugeParams}.

\begin{table}
\begin{center}
\begin{tabular}{|c|c|c|c|c|}
\hline \hline
$\xi^{T}$& $\beta$ & $V$ ($N_s^3 \times N_t$) & $\gamma^*_g$ & $u_s$ \\
\hline \hline
$1$ & $ 6.0 $   & $ 16^3 \times 16 $ & $1$        & $0.8780$ \\
$1$ & $ 6.0 $   & $ 16^3 \times 48 (\dagger) $ & $1$    & $0.8780$ \\
$3$ & $ 6.1 $   & $ 16^3 \times 48 $ & $2.464$    & $0.8279$\\
$6$ & $ 6.1 $   & $ 16^3 \times 96 $ & $4.7172$   & $0.8195$ \\
\hline
\end{tabular} 
\end{center}
\caption{\label{tab:GaugeParams}Parameters of the Quenched Ensembles
  used in this study. In the first column $\xi^{T}$ refers to the desired
  target anisotropies, with the final bare gauge anisotropies $\gamma^*_g$
  being shown in the 4th column. Apart from ($\dagger$) the temporal extent
  grows from $N_t=16$ with the anisotropy. ($\dagger$) was used merely for
  checking the pion mass at the isotropic parameter set since the
  $N_t=16$ case was too short for measuring the pion mass. We also
  show our measurements for the spatial tadpole coefficient $u_s$.}

\end{table}

\subsection{Tuning the Fermion Parameters}
In this study, we have opted to use Clover fermions, with tree-level
tadpole improved clover coefficients as defined in
Ref.~\ref{eq:ClovCoeffs}. In the anisotropic cases, we needed to tune the
fermion anisotropy as well as our quark masses to fall within our
desired range.

Our tuning exercise then comprised of choosing trial values, of
$\gamma_f$ for a selection of values for $m_0$, and computing the pion
dispersion relations for each $(m_0, \gamma_f)$ pair. Since
anisotropic tuning is not the main subject of our paper, we were content to
do this very roughly and were satisfied by a renormalized anisotropy
within 10\% of our target.
To compute the dispersion relation, we extracted the ground state
energies ($E_o$) of the pion for several initial momenta $\vec{p}$, by
fitting the pion correlation function:
\begin{equation}
C(t,\vec{p})  = \sum_{\vec{x}}e^{i \vec{x}.\vec{p}} \langle
\mathcal{O}^{\dagger}(\vec{x},t) \mathcal{O}(\vec{0},0) \rangle \ \stackrel{{
  t\rightarrow \infty}}{\rightarrow} \ 2 A e^{-E_0 \frac{N_t}{2}} \cosh \left(
      E_0 ( \frac{N_t}{2} - t ) \right),
\end{equation}
where $\mathcal{O}$ is a suitable pion interpolating operator:
\begin{equation}
\mathcal{O}(\vec{x},t) = \bar{\psi}(\vec{x},t) \Gamma \psi(\vec{x},t).
\end{equation}
We used spatial momenta ranging in magnitude from
  $| \vec{p}^2 |=0$ to $| \vec{p}^2 |=\left(\frac{3}{2\pi
    N_s}\right)^2$. We averaged the correlation function over the
  momenta that resulted in equal values of $| \vec{p}^2 |$.
We used the interpolating operator $\Gamma=\gamma_5$ for the
zero momentum fits, whereas for the finite momentum fits we used
$\Gamma=\gamma_4\gamma_5$ as we found the signal to be
cleaner. 

Once $E_0$ was determined for all values of $| \vec{p} |^2$, 
we fitted them to the dispersion relation formula:
\begin{equation}
E^2 = \frac{1}{\xi^2} p^2 + \hat{m}^2 , 
\end{equation}
where $\hat{m}=E_0(|\vec{p}|^2=0)$, by a straight line fit, to extract $\xi$.
To compute our estimate for the pion mass, we then computed the mass in units
of the spatial lattice spacing: $a_s m_{\rm latt}=\xi \hat{m}$, and then 
converted this number to physical units assuming $a_s=0.1$fm.

The correlation functions themselves were constructed using gaussian
gauge invariant source smearing \cite{Allton:1993wc}, and stout link smearing
\cite{Morningstar:2003gk}.  No smearing was performed at the sink. Our tuning
calculations were carried out using 40-100 configurations for each
$(m_0, \gamma_f)$ pair. We used the bootstrap method to estimate our
errors on the masses and fitted anisotropies with 200 bootstrap
samples in each case. When estimating the physical pion mass, 
we added the bootstrap errors on $\xi$ and $\hat{m}$ in quadrature,
rather than under the bootstrap.
The results of our tuning are shown in table 
\ref{tab:FermionParams} wherein we  show the fitted fermion anisotropies, and
our estimates of the mass of the pion. 


\begin{table}
\begin{tabular}{|c|c|c|c|c|c|c|} \hline
$\xi^{T}$ &  $m_0$ & $\gamma^{*}_f$ & $\xi $ & $E(\vec{p}=0)$ & $m_\pi$ (MeV) & \# configs used  \\
\hline
1               &  -0.359 &    1      &     1    & 0.383(3)   & 766(6) & 100 \\
1               &  -0.379 &    1      &     1    & 0.296(2)   & 597(4) & 100 \\
1               &  -0.392 &    1      &     1    & 0.225(2)   & 450(4) & 59 \\
\hline
3               &  -0.13  &  2.95     &  3.06(4) & 0.105(2) & 641(15) & 46 \\
3               &  -0.132 &  2.96     &  3.08(4) & 0.097(1) & 597(12) & 46 \\
3               &  -0.135 &  2.96     &  3.03(4) & 0.082(2) & 498(15) & 46 \\
\hline
6               &  -0.058 &   5.43    &  5.99(10) & 0.0578(9) & 693(16) & 50 \\
6               &  -0.061 &   5.63    &  5.96(10) & 0.042(1) & 504(17) & 41 \\
\hline
\end{tabular}
\caption{\label{tab:FermionParams}At each of our target anisotropies $\xi^{T}$,
  we show the bare values of $m_0$, our tuned bare fermion anisotropy
  $\gamma^{*}_f$, and the resulting measured renormalized anisotropies
  $\xi$ as determined from the pion dispersion relation. We also
  show the resulting pion masses in lattice units ($E(\vec{p}=0)$) and
  in physical units ($m_\pi$). The $\xi=1.0$ results were determined
  on the $16^3 \times 48$ lattice.}
\end{table}

\subsection{Results}
With the tuned fermion parameters in hand, we computed the condition
numbers for the various kinds of preconditionings in our three
quenched ensembles. We did not make use of the cutoff
trick in our temporally preconditioned operators.
This time in the isotropic case, we used the $N_t=16$
ensemble, since we did not need long time extents for fitting.
Specifically we computed condition numbers for the unpreconditioned (Unprec), 
and the 4D Schur even-odd preconditioned operator (4D Schur) to use 
as a standard to compare against, as well as the temporally preconditioned
operators combined with both 3D ILU even-odd preconditioning (TPrec+ILU) 
and Schur style preconditioning in 3-dimensions (Tprec+3D Schur). We used 19
configurations from each ensemble to measure the condition numbers.

\begin{figure}
\includegraphics[ width=4in,keepaspectratio=yes]{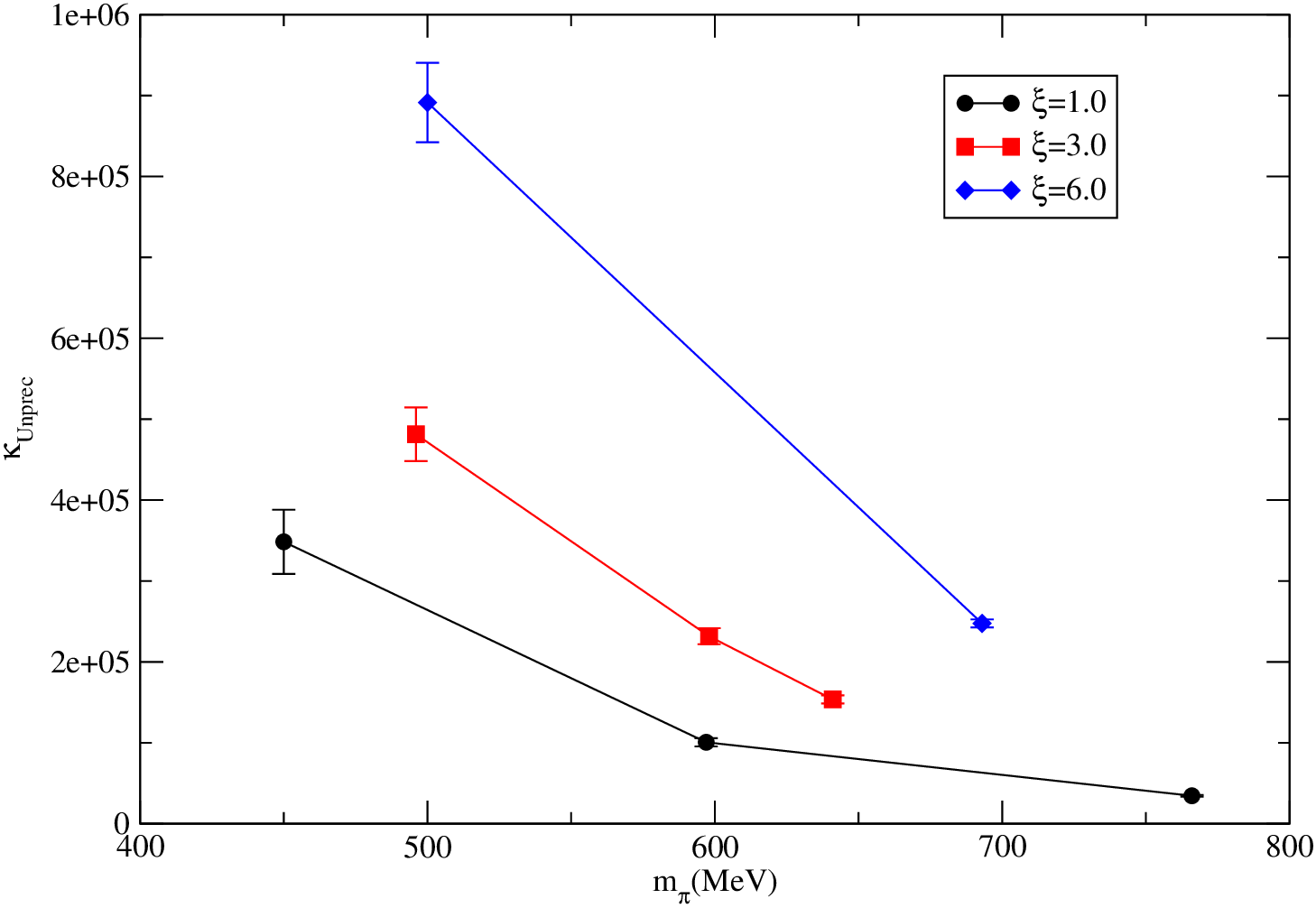}
\caption{\label{fig:Unprec}The condition numbers of the unpreconditioned clover operator against the measured pion mass, for all three of our anisotropy values: $\xi=1.0$ (black circles), $\xi=3.0$ (red squares) and $\xi=6.0$ (blue diamonds).}
\end{figure}

In figure \ref{fig:Unprec} we show how the condition number of the
unpreconditioned operator varies with pion (quark) mass and anisotropy.
We can see, as one would expect, that the condition numbers increase
with decreasing pion (quark) mass as well as with increasing anisotropy.
\begin{figure}
\includegraphics[ width=4in,keepaspectratio=yes]{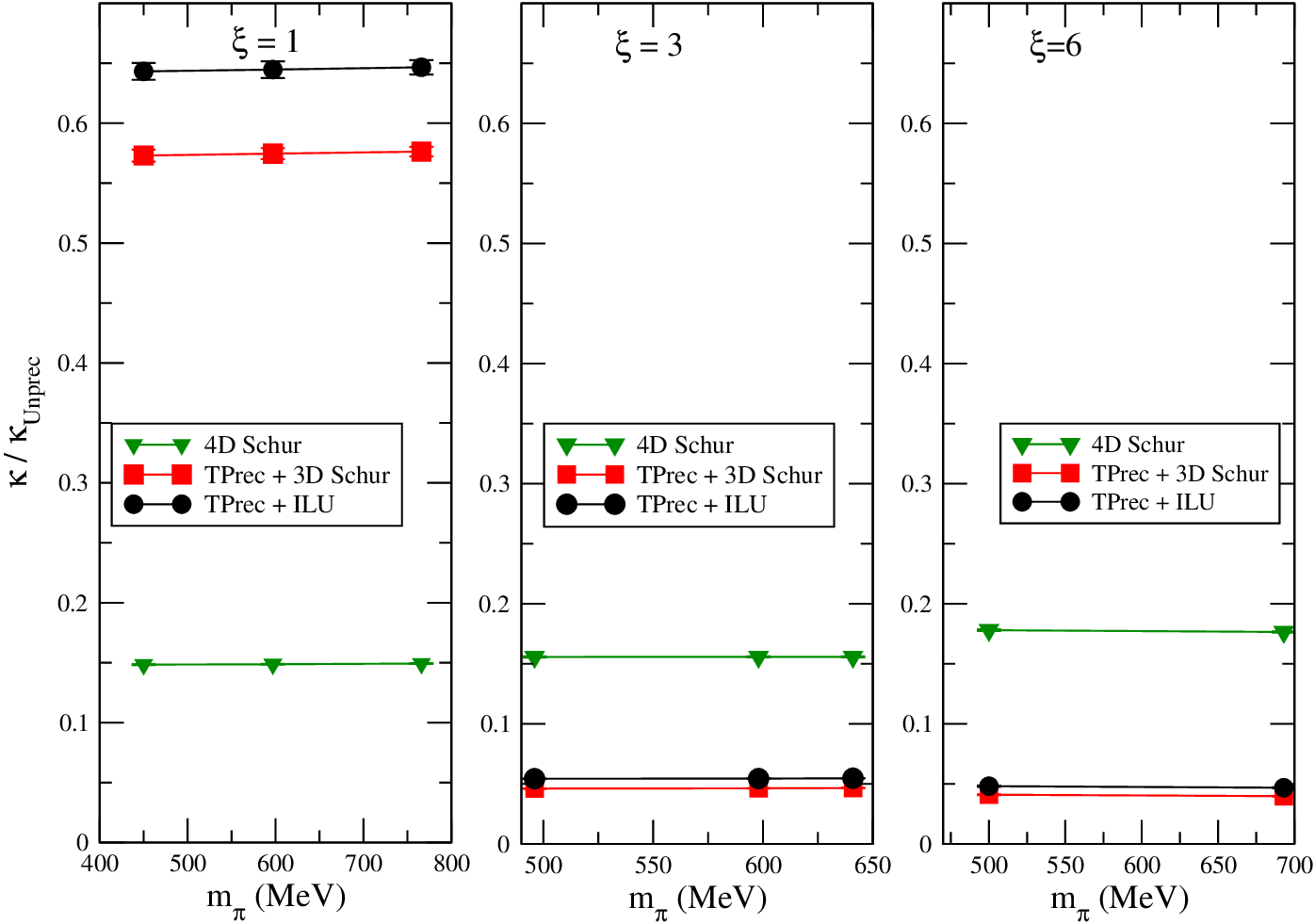}
\caption{\label{fig:Ratios}Ratios of the condition numbers of the
  preconditioned operators to the unpreconditioned operator at the
  same mass and anisotropy (lower is better conditioned). The 3 panes
  are for anisotropies of $\xi=1$ (left), $\xi=3$ (middle), $\xi=6$
  (right). The colors are 4D Schur preconditioning (green triangles),
  temporal preconditioning + 3D ILU (black circles), temporal
  preconditioning + 3D Schur preconditioning (red squares)}
\end{figure}
In figure \ref{fig:Ratios} we plot the ratios of the condition numbers
of the preconditioned operators to the condition number of the
unpreconditioned operator. We separate the results into three graphs
for the three values of $\xi$ used. Separation
in terms of $\gamma_f$ is not practical since for different masses,
one can have different values of $\gamma_f$ for the same renormalized
anisotropy $\xi$.  The ratios give a nice clean signal, as it
appears that from configuration to configuration, the ratios do not
fluctuate very much (although the condition numbers themselves
do). Hence the average of the ratios is very stable.

First we can see in the leftmost graph of fig. \ref{fig:Ratios}, that
in the isotropic case, the greatest gain comes from the 4D Schur
preconditioning, whose condition number is about 15\% of the
unpreconditioned (Unprec.) case. Some gain over the unpreconditioned
case can still be achieved using temporal preconditioning combined
with either 3D Schur (TPrec + 3D Schur) or partial temporal
preconditioning with ILU in 3D (TPrec + ILU). However the temporally
preconditioned cases are not as efficacious as the 4D Schur
preconditioning in the isotropic case.
Looking at the middle and rightmost graphs of fig. \ref{fig:Ratios} we see
that in the anisotropic cases, the temporally preconditioned operators 
fare much better, with condition numbers that are around 4\% - 6\%
of the upreconditioned one.

We note that the mass dependence in these ratios appears to be very
mild, for a given value of $\xi$. Also with the temporal
preconditioning we see an improvement in the condition number ratios as
$\xi$ is increased.  This is presumably due to the suppression factors
of $\frac{1}{\gamma_f}$ and $\frac{1}{\gamma^2_f}$ in the temporally
preconditioned operators. We expect this is because of the clover term
in the preconditioner which has components in both the spatial and
temporal directions, and the temporal components will counteract the
suppression factors $\gamma_f$ and $\gamma_f^2$ to some degree.  It is
however encouraging to see that the partial preconditioning combined
with 3D-ILU even-odd preconditioning in space is nearly as good as the
3D Schur preconditioned case.  This suggests that not dealing with the
clover term in the preconditioner is not in fact catastrophic.  This
is welcome news as the 3D-ILU preconditioning is considerably simpler
to implement and involves fewer FLOPs than the fully preconditioned
3D-Schur approach.

\begin{figure}
\includegraphics[ width=4in,keepaspectratio=yes]{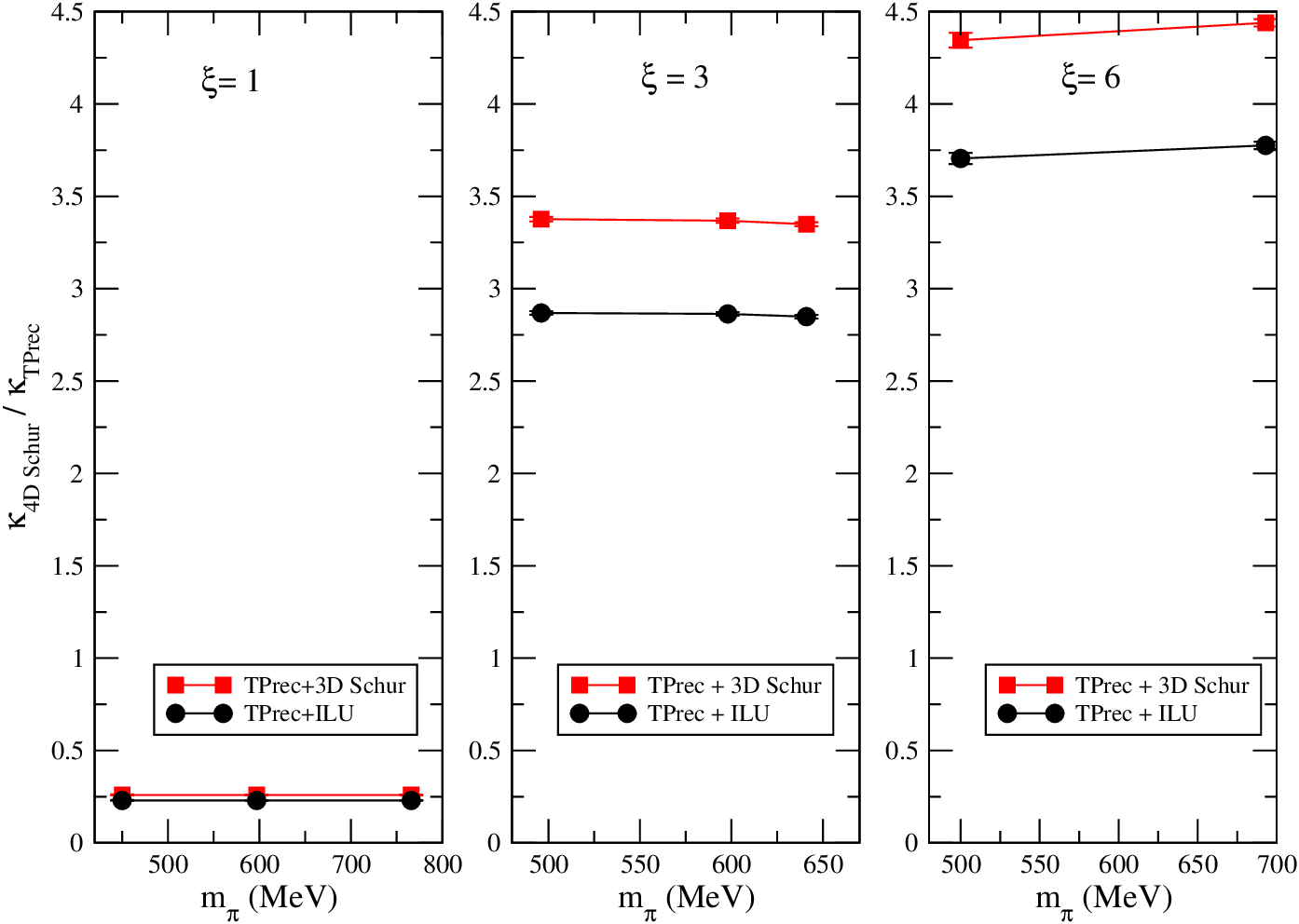}
\caption{\label{fig:Ratios4D}Ratios of the condition numbers of the 4D
  Schur even-odd preconditioned operator to the two temporally
  preconditioned cases (higher is better conditioned). The 3 panes are
  for anisotropies of $\xi=1$ (left), $\xi=3$ (middle), $\xi=6$
  (right). The colors are 4D Schur/TPrec + ILU (black circles), and 4D
  Schur/TPrec + 3D Schur (red squares)}
\end{figure}

We replot the condition number data in figure \ref{fig:Ratios4D}, this
time plotting the ratio of the condition number of the 4D Schur
operator to those of the temporally preconditioned operators, to see
if we gain in condition with respect to the ``standard
preconditioning''.  The ratios are are defined as
\begin{equation}
\frac{\kappa_{\mbox{4D Schur}}}{\kappa_{\mbox{TPrec.+ILU}}}, \ \mbox{and} \frac{\kappa_{\mbox{4D Schur}}}{\kappa_{\mbox{TPrec.+3D Schur}}} \ .
\end{equation}
and hence values larger than one indicates that the temporally preconditioned
operator is better conditioned than the Schur4D, whereas values less than 
one indicate that the Schur4D is better conditioned.

We see, as before, in the leftmost graph of fig. \ref{fig:Ratios4D}, that
in the isotropic case, neither of the temporal preconditioned
approaches can beat the 4D Schur preconditioned approach.  The
condition numbers of the temporally preconditiond operators are about
4 times the Schur 4D case. Looking at the middle and rightmost graphs,
one sees that the TPrec.+ILU preconditinong gives a decrease in the
condition number of about a factor 2.8-3.8 (depending on $\xi$), while
the TPrec+3D Schur preconditioning gives a decrease of a factor of
about 3.3-4.4 compared to the standard 4D Schur even odd
preconditioning (again depending on $\xi$).

At this point, we should recall that the numerical overhead of the
ILU style preconditioning is between 44\%-61\% depending on
implementation. Using the most conservative overhead of 61\%, this
gives an overall gain in {\em total cost} for this scheme, compared to
the 4D Schur case, of $2.8/1.61 - 3.8/1.61$ which equates to between
$1.74-2.36$ for anisotropies of $\xi=3-6$. With the most optimistic
overhead estimate (44\%) these gains can grow to $1.94-2.64$.

\section{Conclusions} \label{s:Conclusions}
  We have demonstrated the technique of temporal preconditioning for anisotropic
discretizations of the Wilson and SW Dirac operator in lattice QCD.  We
discussed the implementation of the technique in an efficient manner, and its
combination with further even-odd preconditioning techniques, in
particular the 3D-ILU and 3D Schur even-odd approaches. 

In the case of Wilson fermions, the 3D-ILU and 3D-Schur approaches
are in fact identical and we have shown that the only terms in the
preconditioned matrix that differ from the identity are suppressed
by two powers of the anisotropy, $\xi^2$.
For SW fermions, the presence of the clover term complicates the implementation
since there are explicit terms that are diagonal in spatial indices and that 
connect the null spaces of the forward and backward projectors. 
The 3D-ILU approach
remains straightforward, however, the resulting preconditioned matrix
has non-zero off diagonal elements, which are only suppressed by factors of 
$\xi$ rather than $\xi^2$. 

We implemented the method in Chroma, a general code base for lattice QCD. 
We have also considered
optimization techniques suggested from work in the realm of GPGPUs.
The appendix shows that the number
of floating point operations in the temporally preconditioned operator
with ILU even-odd preconditioning is about 30\% and 61\% higher than
the corresponding 4D Schur preconditioned operator for the Wilson and
SW operators respectively in the most generic case. Specialized
implementation can reduce floating point the overhad of the Clover action to about
44\%.  These costs may further be ameliorated as needed through
judicious use of the cutoff trick. We have not used the cutoff trick
in the numerical results of this paper.

We have carried out numerical tests in a variety of quenched
ensembles, both isotropic and anisotropic to investigate the efficacy
of the techniques discussed, using Wilson-Clover fermions at a range
of quark masses.

Our chief conclusion is that the technique works well for anisotropic
cases. In our studies, with anisotropies of $\xi=3$ and $\xi=6$, the
temporally preconditioned Clover operator with ILU preconditioning had
condition numbers between a factor of about 2.8-3.3 times smaller than
the usual 4D Schur preconditioned case, whereas the temporally
preconditioned operator with 3D Schur even-odd preconditioning had
condition numbers that were about 3.3-4.4 times smaller than the usual
4D Schur preconditioned case. Combined with the various overhead
estimates, this gives the temporally preconditioned, ILU
preconditioned clover operator a cost advantage of a factor of
$1.74-2.36$ over the more standard 4D Schur preconditioned approach
depending on implementation. The 4D Schur preconditioner, however,
appeared to be the best conditioned in the isotropic case.
 
Due to the decreases in condition number observed using the temporal
preconditioned methods, it becomes attractive to extend the
preconditioning scheme to Hybrid Molecular Dynamics-Monte Carlo
algorithm such as Hybrid Monte Carlo \cite{Duane:1987de}. New terms
will arise in the Hamiltonian, due to the determinants of the
preconditioning matrices.  We leave the full discussion of these
ramifications to a future publication.

\section{Acknowledgements}
This work was done using the Chroma software
suite~\cite{Edwards:2004sx} on clusters at Jefferson Laboratory using
time awarded under the USQCD Initiative.  MP was supported by Science
Foundation Ireland under research grants 04/BRG/P0266 and 07/RFP/PHYF168.  
MP is grateful for the generous
hospitality of the theory center at TJNAF during which time some of this
research was carried out. We would like to thank Mike Clark and Ron Babich
for insightful discussions and pointing out the utility of the Dirac basis
and the Axial gauge to this technique. Authored by Jefferson Science Associates,
LLC under U.S. DOE Contract No. DE-AC05-06OR23177. The U.S. Government
retains a non-exclusive, paid-up, irrevocable, world-wide license to
publish or reproduce this manuscript for U.S. Government purposes.

\bibliographystyle{apsrev}
\bibliography{temp_prec}

\begin{thebibliography}{19}
\expandafter\ifx\csname natexlab\endcsname\relax\def\natexlab#1{#1}\fi
\expandafter\ifx\csname bibnamefont\endcsname\relax
  \def\bibnamefont#1{#1}\fi
\expandafter\ifx\csname bibfnamefont\endcsname\relax
  \def\bibfnamefont#1{#1}\fi
\expandafter\ifx\csname citenamefont\endcsname\relax
  \def\citenamefont#1{#1}\fi
\expandafter\ifx\csname url\endcsname\relax
  \def\url#1{\texttt{#1}}\fi
\expandafter\ifx\csname urlprefix\endcsname\relax\def\urlprefix{URL }\fi
\providecommand{\bibinfo}[2]{#2}
\providecommand{\eprint}[2][]{\url{#2}}

\bibitem[{\citenamefont{Morningstar and Peardon}(1997)}]{Morningstar:1997ff}
\bibinfo{author}{\bibfnamefont{C.~J.} \bibnamefont{Morningstar}}
  \bibnamefont{and} \bibinfo{author}{\bibfnamefont{M.~J.}
  \bibnamefont{Peardon}}, \bibinfo{journal}{Phys. Rev.}
  \textbf{\bibinfo{volume}{D56}}, \bibinfo{pages}{4043} (\bibinfo{year}{1997}),
  \eprint{hep-lat/9704011}.

\bibitem[{\citenamefont{Umeda et~al.}(2003)}]{Umeda:2003pj}
\bibinfo{author}{\bibfnamefont{T.}~\bibnamefont{Umeda}} \bibnamefont{et~al.}
  (\bibinfo{collaboration}{CP-PACS}), \bibinfo{journal}{Phys. Rev.}
  \textbf{\bibinfo{volume}{D68}}, \bibinfo{pages}{034503}
  (\bibinfo{year}{2003}), \eprint{hep-lat/0302024}.

\bibitem[{\citenamefont{Morrin et~al.}(2006)\citenamefont{Morrin, Cais,
  Peardon, Ryan, and Skullerud}}]{Morrin:2006tf}
\bibinfo{author}{\bibfnamefont{R.}~\bibnamefont{Morrin}},
  \bibinfo{author}{\bibfnamefont{A.~O.} \bibnamefont{Cais}},
  \bibinfo{author}{\bibfnamefont{M.}~\bibnamefont{Peardon}},
  \bibinfo{author}{\bibfnamefont{S.~M.} \bibnamefont{Ryan}}, \bibnamefont{and}
  \bibinfo{author}{\bibfnamefont{J.-I.} \bibnamefont{Skullerud}},
  \bibinfo{journal}{Phys. Rev.} \textbf{\bibinfo{volume}{D74}},
  \bibinfo{pages}{014505} (\bibinfo{year}{2006}), \eprint{hep-lat/0604021}.

\bibitem[{\citenamefont{Lin et~al.}(2009)}]{Lin:2008pr}
\bibinfo{author}{\bibfnamefont{H.-W.} \bibnamefont{Lin}} \bibnamefont{et~al.}
  (\bibinfo{collaboration}{Hadron Spectrum}), \bibinfo{journal}{Phys. Rev.}
  \textbf{\bibinfo{volume}{D79}}, \bibinfo{pages}{034502}
  (\bibinfo{year}{2009}), \eprint{0810.3588}.

\bibitem[{\citenamefont{Edwards et~al.}(2008)\citenamefont{Edwards, Joo, and
  Lin}}]{Edwards:2008ja}
\bibinfo{author}{\bibfnamefont{R.~G.} \bibnamefont{Edwards}},
  \bibinfo{author}{\bibfnamefont{B.}~\bibnamefont{Joo}}, \bibnamefont{and}
  \bibinfo{author}{\bibfnamefont{H.-W.} \bibnamefont{Lin}}
  (\bibinfo{year}{2008}), \eprint{0803.3960}.

\bibitem[{\citenamefont{Lin et~al.}(2007)\citenamefont{Lin, Edwards, and
  Joo}}]{Lin:2007yf}
\bibinfo{author}{\bibfnamefont{H.-W.} \bibnamefont{Lin}},
  \bibinfo{author}{\bibfnamefont{R.~G.} \bibnamefont{Edwards}},
  \bibnamefont{and} \bibinfo{author}{\bibfnamefont{B.}~\bibnamefont{Joo}}
  (\bibinfo{year}{2007}), \eprint{arXiv:0709.4680 [hep-lat]}.

\bibitem[{\citenamefont{Sheikholeslami and
  Wohlert}(1985)}]{Sheikholeslami:1985ij}
\bibinfo{author}{\bibfnamefont{B.}~\bibnamefont{Sheikholeslami}}
  \bibnamefont{and} \bibinfo{author}{\bibfnamefont{R.}~\bibnamefont{Wohlert}},
  \bibinfo{journal}{Nucl. Phys.} \textbf{\bibinfo{volume}{B259}},
  \bibinfo{pages}{572} (\bibinfo{year}{1985}).

\bibitem[{\citenamefont{Sherman and Morrison}(1949)}]{Sherman:1949}
\bibinfo{author}{\bibfnamefont{J.}~\bibnamefont{Sherman}} \bibnamefont{and}
  \bibinfo{author}{\bibfnamefont{W.~J.} \bibnamefont{Morrison}},
  \bibinfo{journal}{Annals of Mathematical Statistics}
  \textbf{\bibinfo{volume}{20}}, \bibinfo{pages}{621} (\bibinfo{year}{1949}).

\bibitem[{\citenamefont{Sherman and Morrison}(1950)}]{Sherman:1950}
\bibinfo{author}{\bibfnamefont{J.}~\bibnamefont{Sherman}} \bibnamefont{and}
  \bibinfo{author}{\bibfnamefont{W.~J.} \bibnamefont{Morrison}},
  \bibinfo{journal}{Annals of Mathemetical Statistics}
  \textbf{\bibinfo{volume}{21}}, \bibinfo{pages}{124} (\bibinfo{year}{1950}).

\bibitem[{\citenamefont{Woodbury}(1950)}]{Woodbury:1950}
\bibinfo{author}{\bibfnamefont{M.~A.} \bibnamefont{Woodbury}},
  \bibinfo{journal}{Princeton University, Princeton, N. J., Statistical
  Research Group, Memo. Rep} \textbf{\bibinfo{volume}{42}},
  \bibinfo{pages}{4pp} (\bibinfo{year}{1950}).

\bibitem[{\citenamefont{Edwards and Joo}(2005)}]{Edwards:2004sx}
\bibinfo{author}{\bibfnamefont{R.~G.} \bibnamefont{Edwards}} \bibnamefont{and}
  \bibinfo{author}{\bibfnamefont{B.}~\bibnamefont{Joo}}
  (\bibinfo{collaboration}{SciDAC}), \bibinfo{journal}{Nucl. Phys. Proc.
  Suppl.} \textbf{\bibinfo{volume}{140}}, \bibinfo{pages}{832}
  (\bibinfo{year}{2005}), \eprint{hep-lat/0409003}.

\bibitem[{\citenamefont{Clark et~al.}()\citenamefont{Clark, Babich, Barros,
  Brower, and Rebbi}}]{MikeGPU}
\bibinfo{author}{\bibfnamefont{M.~A.} \bibnamefont{Clark}},
  \bibinfo{author}{\bibfnamefont{R.}~\bibnamefont{Babich}},
  \bibinfo{author}{\bibfnamefont{K.}~\bibnamefont{Barros}},
  \bibinfo{author}{\bibfnamefont{R.}~\bibnamefont{Brower}}, \bibnamefont{and}
  \bibinfo{author}{\bibfnamefont{C.}~\bibnamefont{Rebbi}}, \bibinfo{note}{in
  preparation}.

\bibitem[{\citenamefont{Barros et~al.}(2008)\citenamefont{Barros, Babich,
  Brower, Clark, and Rebbi}}]{MikeGPU2}
\bibinfo{author}{\bibfnamefont{K.}~\bibnamefont{Barros}},
  \bibinfo{author}{\bibfnamefont{R.}~\bibnamefont{Babich}},
  \bibinfo{author}{\bibfnamefont{R.}~\bibnamefont{Brower}},
  \bibinfo{author}{\bibfnamefont{M.~A.} \bibnamefont{Clark}}, \bibnamefont{and}
  \bibinfo{author}{\bibfnamefont{C.}~\bibnamefont{Rebbi}},
  \bibinfo{journal}{POS} \textbf{\bibinfo{volume}{LATTICE200}},
  \bibinfo{pages}{045} (\bibinfo{year}{2008}),
  \urlprefix\url{http://www.citebase.org/abstract?id=oai:arXiv.org:0810.5365}.

\bibitem[{\citenamefont{Bunk}(1997)}]{Bunk:1996kt}
\bibinfo{author}{\bibfnamefont{B.}~\bibnamefont{Bunk}}, \bibinfo{journal}{Nucl.
  Phys. Proc. Suppl.} \textbf{\bibinfo{volume}{53}}, \bibinfo{pages}{987}
  (\bibinfo{year}{1997}), \eprint{hep-lat/9608109}.

\bibitem[{\citenamefont{Michael and Shanahan}(1996)}]{Michael:1995bi}
\bibinfo{author}{\bibfnamefont{C.}~\bibnamefont{Michael}} \bibnamefont{and}
  \bibinfo{author}{\bibfnamefont{H.}~\bibnamefont{Shanahan}}
  (\bibinfo{collaboration}{UKQCD}), \bibinfo{journal}{Nucl. Phys. Proc. Suppl.}
  \textbf{\bibinfo{volume}{47}}, \bibinfo{pages}{337} (\bibinfo{year}{1996}),
  \eprint{hep-lat/9509083}.

\bibitem[{\citenamefont{Klassen}(1998)}]{Klassen:1998ua}
\bibinfo{author}{\bibfnamefont{T.~R.} \bibnamefont{Klassen}},
  \bibinfo{journal}{Nucl. Phys.} \textbf{\bibinfo{volume}{B533}},
  \bibinfo{pages}{557} (\bibinfo{year}{1998}), \eprint{hep-lat/9803010}.

\bibitem[{\citenamefont{Allton et~al.}(1993)}]{Allton:1993wc}
\bibinfo{author}{\bibfnamefont{C.~R.} \bibnamefont{Allton}}
  \bibnamefont{et~al.} (\bibinfo{collaboration}{UKQCD}),
  \bibinfo{journal}{Phys. Rev.} \textbf{\bibinfo{volume}{D47}},
  \bibinfo{pages}{5128} (\bibinfo{year}{1993}), \eprint{hep-lat/9303009}.

\bibitem[{\citenamefont{Morningstar and Peardon}(2004)}]{Morningstar:2003gk}
\bibinfo{author}{\bibfnamefont{C.}~\bibnamefont{Morningstar}} \bibnamefont{and}
  \bibinfo{author}{\bibfnamefont{M.~J.} \bibnamefont{Peardon}},
  \bibinfo{journal}{Phys. Rev.} \textbf{\bibinfo{volume}{D69}},
  \bibinfo{pages}{054501} (\bibinfo{year}{2004}), \eprint{hep-lat/0311018}.

\bibitem[{\citenamefont{Duane et~al.}(1987)\citenamefont{Duane, Kennedy,
  Pendleton, and Roweth}}]{Duane:1987de}
\bibinfo{author}{\bibfnamefont{S.}~\bibnamefont{Duane}},
  \bibinfo{author}{\bibfnamefont{A.~D.} \bibnamefont{Kennedy}},
  \bibinfo{author}{\bibfnamefont{B.~J.} \bibnamefont{Pendleton}},
  \bibnamefont{and} \bibinfo{author}{\bibfnamefont{D.}~\bibnamefont{Roweth}},
  \bibinfo{journal}{Phys. Lett.} \textbf{\bibinfo{volume}{B195}},
  \bibinfo{pages}{216} (\bibinfo{year}{1987}).

\end{thebibliography}

\appendix
\subsection{Appendix -- Full Preconditioning with Clover}\label{s:FullClover}
In this appendix, we consider the question of
how to deal with the inversion of the term  $\left(A + \mu - D_t \right)$,
in the case of full temporal preconditioning.

We begin with 
\begin{equation}
A(\vec{x},t) + \mu - D_t(\vec{x},t) =  A(\vec{x},t) + \mu 
 - P_{-} U_{t}(\vec{x},t)\delta_{\vec{x},t+1;\vec{x},t'}
              - P_{+}  U^{\dagger}_{t}(\vec{x},t-1)\delta_{\vec{x},t-1; \vec{x},t}
\end{equation}

Inverting $A+\mu-D_t$ can be done with a single step Woodbury procedure:
\begin{equation}
A + \mu - D_t = T_0 + V W^\dagger
\end{equation}
where
\begin{equation}
V = \left( \begin{array}{c}
-U^\dagger(\vec{x}, N_t-1) P_{+} \\
0 \\
0 \\
\vdots \\
0 \\
-U(\vec{x}, N_t-1) P_{-}
\end{array} \right) \quad \mbox{,}\quad W^\dagger = \left( P_{-} , 0, 0, 0, \ldots P_{+} \right)
\end{equation}

With the tridiagonal matrix $T_0$ is now, supressing space indices:
\begin{equation}
T_0 = \left( \begin{array}{ccccc} 
A(0) + \mu  & -U( 0) P_{-}& 0 & \ldots & 0 \\
-U^\dagger( 0) P_{+} & A(1)+\mu &  -U( 1)P_{-} & 0 & \ldots \\
 0       & \ddots & \ddots & \ddots & \vdots \\
 \vdots  &        &  -U^\dagger( N_t-3) P_{+}      &  A( N_t -2 )+\mu   &  -U( N_t -2 )P_{-} \\
0  &  0 & \vdots      &  -U^\dagger( N_t-2) P_{+}      &  A( N_t - 1 )+\mu
\end{array}
\right) 
\end{equation}

This matrix, while easy to apply, is not immediately straighforward to
invert, because of its projector structure. In our numerical work, to
gauge the efficacy of this approach, we used an inner conjugate
gradients algorithm to invert this matrix.

\subsection{Floating Point Operation count for ILU Scheme}\label{s:Count}

Let us recount the the number of floating point operations needed to apply the usual 4D Schur preconditioned operator:
\begin{equation}
\tilde{M}_{{\rm 4D Schur}} = A^{oo} - \frac{1}{4} D^{oe} \left(A^{-1}\right)^{ee} D^{eo}
\end{equation}
where the mass term has been absorbed into the diagonal part of $A$
and $D^{oe}$ denotes the 4D Hopping matrix. The parameters $\gamma_f$ have been absorbed into a rescaling of the gauge links. We will
use the notation $\C(Q)$ do denote the floating point cost of some
generic term $Q$.  Applying the $A^{oo}$ term and $A^{ee}$ both
require 522 FLOPs per site, on $V/2$ sites (single checkerboard) each,
giving $\C(A^{oo}) = \C(A^{ee})=522(V/2)$ FLOPs

The $D^{oe}$ and $D^{eo}$ terms require 1320 FLOPs each per site on
$V/2$ sites (single checkerboard) each. This number is arrived at by
considering the spin projection operators as having no floating point operations, since only sign flips and exchanges of real and imaginary components are involved. Each $SU(3)$ matrix-vector (matvec) operation requires $66$ FLOPs,
corresponding to three inner products between the three rows of the matrix, and the column vector. Each inner product involves 3 complex multiplies (6 FLOPs each) and 2 complex adds (2 FLOPs each), or $3 \times 6 + 2 \times 2 = 22$ FLOPs, giving a total of $3 \times 22= 66$ for the three inner products in the complete matvec operation.

Spin reconstuction (recons) takes $12$ FLOPs per site, coming from a single complex add for each of 2 spin-color components (6 complex-adds in total); again, not counting sign flips and real complex component interchanges. Finally, $24$ FLOPs are required to sum two (now reconstructed) 4 spinors (sumvec4) to evaluate the sum over directions (4 spin-color components, so 12 complex components in total, with 2 FLOPs per component).

In $N_d$ dimensions 
\begin{equation}
\C( D^{oe} ) = 2 N_d(N^{*}_\gamma{\rm matvec}+{\rm recons})+ {\rm sumvec4}\times(2N_d-1)
\end{equation}
where the factor of $2 N_d$ comes from doing forward and backward
projections and reconstructions in $N_d$ dimensions, and $N^{*}_\gamma$
is the number of spin components left after spin projection.
So, in 4-dimensions, with $N_\gamma=4$ we have $N_d=4, N^{*}_\gamma=2$
and one has:
\begin{equation}
\C( D^{oe} ) = 8 ( 2 \times 66 + 12 )+ 24 \times 7 = 1320 \ {\rm FLOPs}
\end{equation}

Correspondingly, per site the 4D Schur Operator takes up 
\begin{equation}
\C( \tilde{M}_{\rm 4D Schur} ) = 2 \times \left(1320+522 \right) + 48 = 3732  \ {\rm FLOPs}
\end{equation}
where the last factor of $48$ comes from the AXPY operation to apply the factor of $\frac{1}{4}$ and subtracting the two terms from each other. Hence, applying the operator costs  $3732(V/2)$ FLOPs in total, which it is convenient to re-express as $3732 N_t (V_s/2)$, with $N_t$ the extent of the time direction, and $V_s$ being the number of spatial coordinates per timeslice.

Let us now consider the ILU preconditioned operator
\begin{equation}
\tilde{M}_{\rm ILU} = \left( \begin{array}{cc} 
1 + \bar{A}^{ee} & \bar{A}^{ee} \bar{D}^{eo}_s \\
\bar{D}^{oe} \bar{A}^{ee} &  1 + \bar{A}^{oo} - \bar{D}^{oe}_s \left[ 1 - \bar{A}^{ee} \right] \bar{D}^{eo}_s \end{array}\right)
\end{equation}
where we have absorbed the factors of $\xi$ into the $\bar{D}$ terms.
Applying $\tilde{M}_{\rm ILU}$ to some vector $\psi$, to result in $\chi = \tilde{M}_{\rm ILU} \psi$ implies that
\begin{eqnarray}
\chi_e &=& \psi_e + \bar{A}^{ee} \left( \bar{D}^{eo} \psi_o + \psi_{e} \right) \\
\chi_o &=& \psi_o + \bar{A}^{oo} \psi_o + \bar{D}^{oe} \left\{ \bar{A}^{ee} \left[ \bar{D}^{eo} \psi_o + \psi_e \right] - \bar{D}^{eo} \psi_o \right\}
\end{eqnarray}
and one can reuse several terms between $\chi_e$ and $\chi_o$. Thus $\tilde{M}_{\rm ILU}$ can be efficiently applied by computing in sequence:
\begin{eqnarray}
\chi_e &=& \psi_e \\
\chi_o &=& \psi_o \\
t_1 &=& \bar{D}^{eo} \psi_{o} \\
t_2 &=& \bar{A}^{ee} \left( t_1 + \psi_e \right) \\
\chi_e &=& \chi_{e} + t_2 \\
\chi_o &=& \chi_o + \bar{A}^{oo} \psi_{o} + \bar{D}^{oe}\left( t_2 - t_1\right)
\end{eqnarray}
and apart from 5 vector additions, we need to apply $\bar{A}$ and $\bar{D}$
twice each (once per checkerboard). Since $\bar{A} = C_L A C_R$ and 
$\bar{D} = C_L D C_R$ we have
\begin{equation}
\C( \tilde{M}_{\rm ILU} ) = 5 \times 24 \times (V/2) + 2 \left[  \C( \bar{D}_s ) + \C( \bar{A} )\right] = 120 ( V/2 ) + 2\left[ 2 \C(C_L) + 2 \C(C_R) + \C(A) + \C(D_s) \right]
\end{equation}
where the $5 \times 24 \times (V/2)$ term is to account for the 5 vector adds, each on a single checkerboard. 

We still have $\C(A)=522$ per site and by substituting $N_d = 3$ into the discussion for $\C(D)$ one finds that applications of the 3D $D_s$ operator cost 984 FLOPs  per site. The preconditioners $C_L$ and $C_R$ have the same cost each namely $\C(C)$ 
and so
\begin{equation}
\C( \tilde{M}_{\rm ILU} ) = 120 ( V/2 ) + 2 \left[ 4 \C(C) + \C(A) + \C(D_s) \right]
 = 120 (V/2) + 8\C(C) + ( 2 \times 522 + 2 \times 984 ) (V/2) = 8\C(C) + 3132 (V/2)
\end{equation}

Applying $C$ requires a back (forward) subsitution for each {\em
  spatial site} of the appropriate checkerboard. We consider the
backsubsitution here, but the working is similar (and the FLOP count
is identical) for the forward substitution.  The back subsitution
needs to be performed on only 2 spin color components, after spin
projections with either $1 + \gamma_0$ or $1 - \gamma_0$ for $C_L$ or
$C_R$, followed by an appropriate reconstruction. After the
backsubsitution the Woodbury procedure involves for each spatial site,
working with length $N_t$ block vectors, a matrix multiplication by
the a pre-computed $SU(3)$ matrix, and an $SU(3)$ subtraction. As
usual we do not count any floating point operations for the projection
part of the spin projection steps. This gives
\begin{equation}
\C(C) = N^*_\gamma \left[ (V_S / 2) \left( \C(B) + \C(W) \right) \right]  +(V/2)( 12 + 48 )
\end{equation}
where $V_s$ denotes the number of spatial sites, $\C(B)$ denotes the cost for the
back substitution, $\C(W)$ denotes the rest of the Woodbury process, the 
$12 (V/2)$ FLOPs come from the spin reconstruction on one chekerboard and 
the $48(V/2)$ FLOPs come from adding the $P_{+}$ and $P_{-}$ terms in $C_L$ and $C_R$ ; one vector addition costing 24 FLOPs and an overall scaling by $\frac{1}{2}$ to normalize the projectors costing another 24 FLOPs.

We now need to consider the backsubstution on a single spin color component:
The first step is just a scaling by the diagonal $\psi_{N_t-1} = \frac{1}{\mu} \chi_{N_t-1}$ or 6 FLOPs. Then there follow $N_t - 1$ steps of $\psi_i = \frac{1}{\mu} \left[ \chi_i - U_i \psi_{i+1} \right]$, each one comprised of an $SU(3)$ matrix vector multiply (66 FLOPs) and a subtraction and a scaling by $\mu$ (6 FLOPs each). In total:
\begin{equation}
\C(B) = 6 + \left(N_t-1\right) \left( 66 + 6 + 6  \right) = 78N_t - 72 \ {\rm FLOPs}
\end{equation}

The remainder of the Woodbury procedure involves computing the $X \Lambda W^\dagger$ term. This can be achieved by precomputing $X \Lambda W^\dagger$ for each value of $t$ at initialization, and then this process costs only $N_t$ matrix vector operations (66FLOPs each), and finally we need to subtract  the result of this matrix multiply from the result of the back/forward substitution (6 FLOPs) for each value of $N_t$ and so
\begin{equation}
\C(W) = (66 + 6) N_t = 72 N_t \ {\rm FLOPs}
\end{equation}
and 
\begin{eqnarray*}
\C(C) &=& N^*_\gamma (V_s/2) \left[ 78N_t - 72 + 72N_t  \right] + (V/2)\left[12 + 48\right]\\
 &=& N^*_\gamma  (V_s/2) \left[ 150N_t - 72 \right] + 60(V/2) \\
 &=&  (V_s/2)\left( 300N_t  - 144  \right) + 60 N_t (V_s/2) \\
 &=& (V_s/2) \left( 360N_t  - 144  \right) \ {\rm FLOPs}
\end{eqnarray*}
where we have used $V_s V_t = V$, and $N^*_\gamma=2$ so
\begin{eqnarray}
\C(\tilde{M}_{\rm ILU}) &=& 8\C(C)+ 3132 (V/2) \nonumber \\
&=& 8 (V_s/2) \left( 360 N_t - 144 \right) + 3132 N_t (V_s/2)\nonumber \\
&=& (V_s/2) \left( 2880 N_t - 1152\right) +  3132 N_t (V_s/2) \nonumber \\
&=& \left( 6012 N_t - 1152 \right) (V_s/2) \ {\rm FLOPs} \label{eq:CloverCost}
\end{eqnarray}

Comparing the costs of the two preconditioned operators we have for Clover 
fermions
\begin{equation}
R = \frac{\C( \tilde{M}_{\rm ILU})}{\C(\tilde{M}_{\rm 4D Schur})} = \frac{ 6012 N_t - 1152  }{3732 N_t  } \approx 1.611 - \frac{0.309}{N_t}
 \end{equation}
and so we consider two limiting cases: in the first instance when $N_t=1$ we have $R=1.302$ to 3 decimal places (d.p.) and when $N_t$ is sufficiently large that the term involving it is negligible we have $R=1.611$ to 3 d.p. In a typical case, when $N_t=128-256$ one has $R\approx1.61$ to 2 d.p.

Similarly we can consider the case for just Wilson fermions. In this case 
\begin{equation}
\tilde{M}_{{\rm 4D Schur}} = \mu - \frac{1}{4\mu} D^{oe} D^{eo}
\end{equation}
and the cost is 
\begin{equation}
\C(\tilde{M}_{{\rm 4D Schur}})  = (48 + 2 \times 1320)N_t (V_s/2) = 2668 N_t (V_s/2 ) \ {\rm FLOPs}
\end{equation}
where the $48$FLOPS comes from the subtraction, and scaling (AXPY) operation.

The temporally preconditioned scheme needs only the evaluation of 
\begin{equation}
\tilde{M}_{{\rm ILU}} = I - \bar{D}^{oe}_s \bar{D}^{eo}
\end{equation}
and so the cost in flops is:
\begin{eqnarray}
\C(\tilde{M}_{\rm ILU}) &=& 24(V/2) + 4\C(C) + 2\C(D_s)  \nonumber \\
  &=& 24 N_t (V_s/2) + 4 (V_s/2) \left( 360N_t  - 144  \right) + 2\times 984 N_t (V_s/2) \nonumber \\
&=& (V_s/2)\left( 24 N_t + 1440 N_t - 576 + 1968 N_t \right) \nonumber \\
&=& (V_s/2)\left( 3432 N_t - 576\right)  \ {\rm FLOPs} \label{eq:WilsonCost}
\end{eqnarray}
and  we have
\begin{equation}
R = \frac{ 3432 N_t - 576}{2668 N_t} \approx 1.286-0.216\frac{1}{N_t}
\end{equation}
giving $R$ in the range of $R \approx 1.07-1.286$. 

Use of the cutoff trick removes the multiplication by components of
$X$ and subtraction of the term $Xq$ in steps 2(b) and 2(c) of the
Sherman-Morrison-Woodbury process. If the cutoff value of $t$ is $k$,
one saves $k$ timeslices; from $t=0$ to $t=k-1$; per spatial site in
the ILU Clover operatorevaluting $T^{-1}$ or
$\left( T^\dagger \right)^{-1}$, on each of $N^*_\gamma=2$ spin
components. On one spin component one saves $66+6$ flops, and so per
spatial site, one saves $2\times (66+6) = 144$ FLOPs per
per timeslice; altogether $144k$ FLOPS. One does this every $C_L$
and $C_R$ of which there are 4 in the Wilson case, and 8 in the ILU
preconditioned Clover case, each of which is evaluated on a single
checkerboard ($V_s/2$ sites). Correspondingly
the cutoff trick on $k$ timeslices saves $576 k (V_s/2)$ FLOPs for
the ILU Wilson and $1152 k (V_s/2)$ FLOPs for the ILU Clover operator.

\subsection{Dirac Basis and Axial Gauge} \label{s:CountWithTricks}

The use of the Dirac basis and the Axial gauge was advocated in \cite{MikeGPU,MikeGPU2}
to save FLOPs and memory bandwidth in the implementations of the Dirac
operator on GPU systems. The use of these techniques is also beneficial
in the case of temporal preconditioning. We analyze below the benefits
for temporal preconditioning in terms of the FLOPs savings, but note
also, that savings in memory bandwidth resulting from the use of these
techniques can also help the performance of the implementations.

The use of the Dirac basis, where $\gamma_0$ is diagonal, can save the
cost of spinor reconstruction in the time direction, saving some 48
FLOPs per site in the 4-dimensional $D$ operators -- 12 each in the
forward and backwards time directions respectively from not having
to reconstruct, and another 12 each when accumulating since now only half
vectors need to be accumulated, rather than the reconstructed 4
vectors.  This corresponds to a saving of $48 N_t (V_s/2)$ FLOPs, per
$D$ operator, or a total of $96 N_t(V_s/2)$ when considering the full
4D Schur preconditioned operator (where $D$ is applied twice).

Correspondingly, this trick can save the $(V/2)(12 + 48)=60N_t (V_s/2)$ FLOPs term from the cost of applying $C_L$ and $C_R$ since no spinor reconstruction is needed and one does not need to expend FLOPs when adding the results of the $P_{+}$ and $P_{-}$ projectors, as the projectors will simply write their results to different components of the final 4 spinor. In the case of temporal preconditioned clover fermions, where the preconditioner is used 8 times overall this results in a saving of $480N_t (V_s/2)$ FLOPs. In the case of unimproved fermions, where the preconditioner is used only 4 times one can save $240N_t(V_s/2)$ FLOPs.

Use of the axial gauge allows one to save further FLOPs.  In this case
all the links in the temporal direction (apart from the boundary) are
transformed to have value $U_i=1$. In the case of the 4D $D$ operator,
this saves 2 SU(3) matrix vector multiplies per spin component, coming
from the forward and backward temporal link matrices on the
non-boundary sites. To make counting easier, we'll assume the number
of boundary sites is negligible and assume the saving for every
site. Hence we count $2 N^{*}_\gamma \times 66=264$ FLOPs saved per
site. Two applications of $D$ are needed for the 4D Schur
preconditioned operators, so this trick saves roughly $264$ FLOPs per
site or $528 N_t(V_s/2)$ FLOPs in total.

Correspondingly the $N_t -1 $  back(forward) substitution steps change from $\psi_i=\frac{1}{\mu}\left[ \chi - U_i \psi_{i+1} \right]$ to merely  $\psi_i=\frac{1}{\mu}\left[ \chi - \psi_{i+1} \right]$ and one saves $N_t -1$ matrix multiplies in applying the preconditioner for each spin component, leading to a cost saving of $66 N^{*}_\gamma (N_t -1) = 132N_t -132$ flops per spatial site.

In the case of Clover fermions, where 8 applications of the preconditioner needed, the saving is $(1056N_t-1056)V_s/2$ FLOPSs, whereas in the case of unimproved Wilson, where 4 applications are needed the saving is $(528N_t-528)(V_s/2)$.

Combining the savings from the use of the Dirac basis and the use of the axial gauge one gets, for the cost of the temporally preconditioned Clover Operator:
\begin{equation}
\mathcal{C}(\tilde{M}_{\rm ILU}) = (4476 N_t - 96)(V_s/2)
\end{equation}
which is a saving of roughly 25\% over the previous cost in eq. \ref{eq:CloverCost}.
The cost of the 4D Schur preconditioned operator also reduced:
\begin{equation}
\mathcal{C}(\tilde{M}_{\rm 4DSchur})= 3108 N_t (V_s/2)
\end{equation}
giving the overhead from the preconditioning to be:
\begin{equation} 
R = 1.440 - \frac{0.031}{N_t}
\end{equation}

In the unimproved Wilson Operator one has
\begin{equation}
\mathcal{C}(\tilde{M}_{\rm ILU}) = (2664 N_t - 48)(V_s/2)
\end{equation}
corresponding to a saving of about 22\% over the previous case in eq. \ref{eq:WilsonCost}.

The cost of the 4D Schur preconditioned operator becomes:
\begin{equation}
\mathcal{C}(\tilde{M}_{\rm 4DSchur})= 2044 N_t (V_s/2)
\end{equation}
giving 
\begin{equation} 
R = 1.30 - \frac{0.023}{N_t} \ . 
\end{equation}

Thus, the overhead of temporal preconditioning is between roughly 41-43\% for Clover, and 27-30\% for the unimproved Wilson Case.

\end{document}